\documentclass[twocolumn,twoside,slac_two]{revtex4}
\usepackage{graphicx}
\usepackage{fancyhdr}
\begin{document}

\title{Searches for Lepton Flavor Violation and Lepton Number 
Violation in Charged Lepton Decays}

%

\author{F. Renga}
\affiliation{Paul Scherrer Institut PSI, CH-5235 Villigen, Switzerland and INFN, Sezione di Roma}

\begin{abstract}
I present here the most recent experimental results in the search for Lepton Flavor and Lepton Number Violating decays of charged leptons,
and the perspectives in this field for the next future.
\end{abstract}

\maketitle

\thispagestyle{fancy}

\section{INTRODUCTION}
Lepton Flavor (LF) violation in charged lepton decays is a standard probe for searches of New Physics (NP) beyond 
the Standard Model (SM). Historically, the negative outcomes of these searches contributed to the formulation of the 
SM, where the conservation of the LF is an accidental symmetry, not related to the gauge structure of the 
model itself, but obtained as a consequence of its particle content. As a result, LF is naturally violated in most of the 
extensions of the SM, and the present limits already put strong constraints on their formulation. Moreover, since unobservable
rates are expected in the SM\footnote{Even if the neutrino oscillations are accounted for, the BR of e.g. $\mu \to e \gamma$ is
predicted to be below $10^{-40}$.}, an observation of LF Violation (LFV) in charged lepton decays would provide an unambiguous 
evidence of NP.

Among the many NP models that would produce observable LFV rates (including theories with extra dimensions, unparticles,
etc.), the most popular example is provided by Supersymmetry (SUSY). In SUSY models, even if the theory is formulated to be flavor
blind at some high energy scale, LF violating terms arise in the slepton mass matrix, through 
renormalization group equations, at the electroweak scale. As a result, relatively large contributions to LFV
in charged lepton decays are almost unavoidable, and in most cases these processes are expected to be 
within the reach of present experiments. For example, SUSY models with Grand Unification (SUSY GUTs) tend to predict
the Branching Ratio (BR) of the $\mu \to e \gamma$ and $\tau \to \mu \gamma$ decays to be at a level of $10^{-14}-10^{-10}$
and $10^{-7}-10^{-11}$ respectively, so that, as we will see, a large portion of the parameter space for these models is 
already ruled out by the present limits.

Another interesting feature of LFV is, in many models, its correlation with a possible deviation of the
anomalous magnetic moment of the muon, $g-2$, from the SM expectations~\cite{gm2}, which constitutes, 
at present, the most relevant tension between electroweak precision measurements and the SM. If such a deviation
is real, and is explained within the framework of SUSY, the models tend to predict a large rate for 
$\mu \to e \gamma$, as shown for instance in~\cite{isidori_gm2}.

\begin{figure}[b]
\centering
\includegraphics[width=60mm]{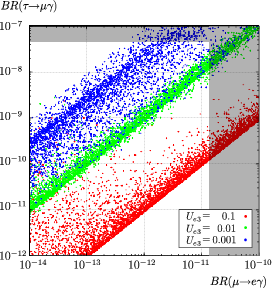}

\vspace{0.5cm}
\includegraphics[width=60mm]{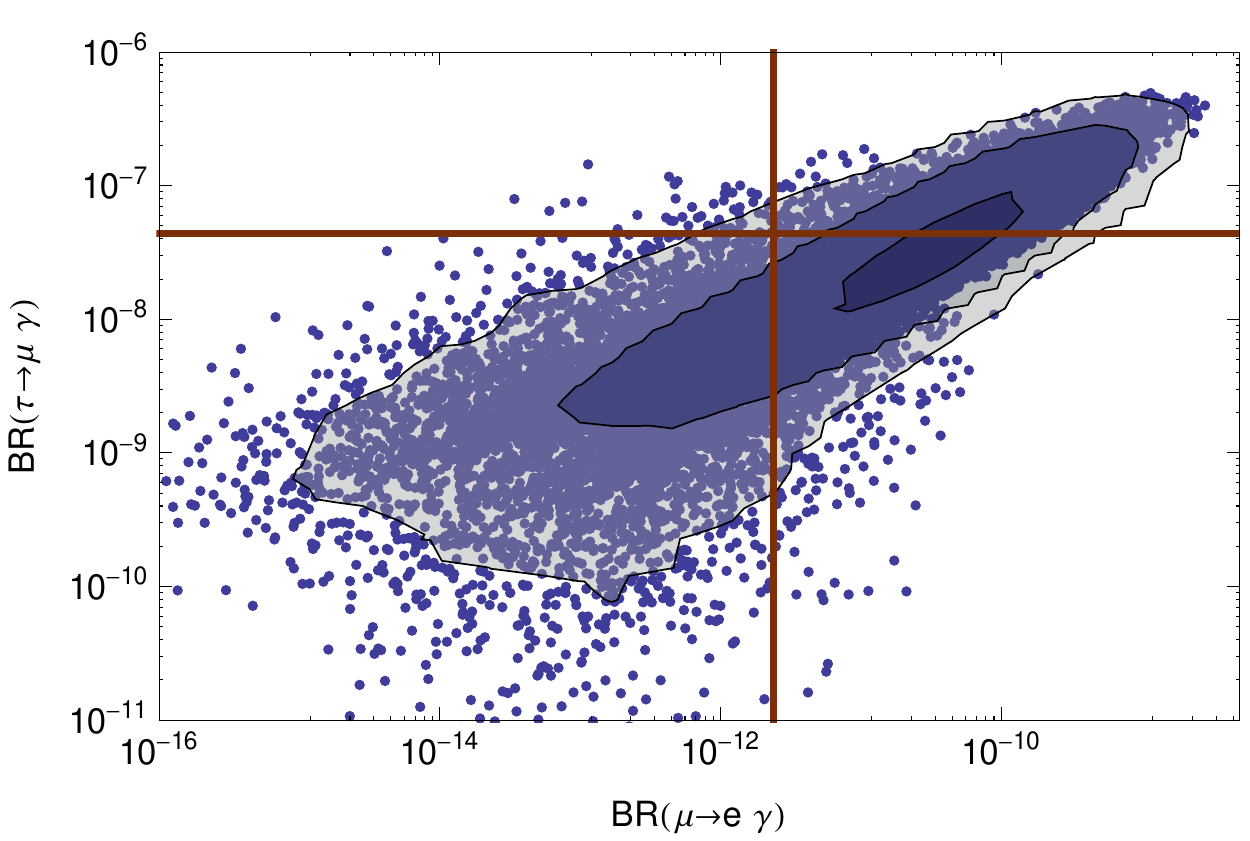}
\caption{Expected $\tau \to \mu \gamma$ and $\mu \to e \gamma$ rates in different SUSY models. Top: SUSY $SU(5)^5$ from~\cite{hisano},
 for different values of $\theta_{13}$. Bottom: SUSY with minimally broken Flavor symmetries and the measured value of $\theta_{13}$ used 
in the calculations, from~\cite{isidori}.} \label{fig:models}
\end{figure}

From the experimental point of view, it is important to stress the high complementarity that exists between the LFV searches in
$\tau$ and muon decays. The point is that the relative rate of LFV decays of $\tau$'s and muons strongly depends on the
specific flavor structure of NP. For most of the SUSY GUTs, large values of the neutrino mixing angle $\theta_{13}$,
like the one recently measured at Daya Bay~\cite{DayaBay}, tend to enhance the muon LFV~\cite{hisano}. Anyway, 
according to other models, it is still possible to have both muon and $\tau$ LFV within the experimental 
reach in the next future~\cite{isidori}. These features are illustrated in Figure~\ref{fig:models} and demonstrate how searches in both
$\tau$ and muon sector are necessary in order to constraint as many models as possible and increase the chances of
discovery.

This presentation covered the most recent results in the search for LFV in charged lepton decays. In the
recent past, the searches in the $\tau$ sector have been mainly carried on at the $B$-Factories PEP-II (SLAC, USA) and KEKB (KEK, Japan), by the BaBar
and Belle experiments, thanks to the large amount of $\tau$ leptons produced in $e^+ e^-$ collisions around 10~GeV~\footnote{The
$e^+e^- \to \tau^+ \tau^-$ cross section at such energies, at the level of 1~nb, is as large as the $e^+ e^- \to b \overline b$
cross section at the $\Upsilon(5S)$ peak.}. 
Now, LHC experiments at CERN are just entering the game, with impressive and in part unexpected performances.  
The muon sector have been instead explored, in the last few years, by the MEG experiment, that searches for the $\mu \to e \gamma$
decay with a sensitivity that reached a few $10^{-12}$. Beside LFV, new results have been recently produced in the search for 
global Lepton Number Violation (LNV), that can be used to constrain models with Majorana neutrinos. These results are also 
covered here.

\section{LFV AND LNV IN THE $\tau$ SECTOR}

In the past decade, the $B$-Factory experiments BaBar and Belle allowed to set the most stringent constraints presently
available for LFV and LNV in $\tau$ decays. Their latest limits for the most popular decay channels $\tau \to \ell \gamma$ and
$\tau \to 3\ell$, summarized in Table~\ref{tab:mu_eg_3l}, are available since a few years, and will not be reviewed in detail here (charge
conjugation is understood in the table and elsewhere through this paper).

\begin{table}[b]
\begin{center}
\caption{Latest results from BaBar and Belle for $\tau \to e \gamma$ and $\tau \to 3\ell$.}
\begin{tabular}{|l|c|c|}
\hline 
\textbf{Channel} & \multicolumn{2}{|c|}{\textbf{90\% C.L. Upper Limit [$\times 10^{-8}$]}} \\
 & \multicolumn{1}{|c}{\textbf{~~~~~~~BaBar~~~~~~~}} & \textbf{~~~~~~~Belle~~~~~~~~} \\
\hline 
$\tau^+ \to e^+ \gamma$ & 3.3~\cite{babar_tlg} & 12~\cite{belle_tlg} \\ 
\hline
$\tau^+ \to \mu^+ \gamma$ & 4.4~\cite{babar_tlg} & 4.5~\cite{belle_tlg} \\ 
\hline
$\tau^+ \to e^+ e^+ e^-$ & 3.4~\cite{babar_t3l} & 2.7~\cite{belle_t3l} \\ 
\hline
$\tau^+ \to e^+ \mu^+ \mu^-$ & 4.6~\cite{babar_t3l} & 2.7~\cite{belle_t3l} \\ 
\hline
$\tau^+ \to e^- \mu^+ \mu^+$ & 2.8~\cite{babar_t3l} & 1.7~\cite{belle_t3l} \\ 
\hline
$\tau^+ \to \mu^+ e^+ e^-$ & 3.7~\cite{babar_t3l} & 1.8~\cite{belle_t3l} \\ 
\hline
$\tau^+ \to \mu^- e^+ e^+$ & 2.2~\cite{babar_t3l} & 1.5~\cite{belle_t3l} \\ 
\hline
$\tau^+ \to \mu^+ \mu^+ \mu^-$ & 4.0~\cite{babar_t3l} & 2.1~\cite{belle_t3l} \\ 
\hline
\end{tabular}
\label{tab:mu_eg_3l}
\end{center}
\end{table}

Instead, we will concentrate our attention to the most recent results, concerning $\tau$
decays with hadrons in the final state. In particular, both Belle and BaBar produced interesting
limits on the $\tau \to \ell h$, $\tau \to \ell h h'$ and $\tau \to \Lambda h$ decays, where
$h$ and $h'$ indicate generic hadrons.

\subsection{Searches for $\tau \to \ell h^0$} 
\label{sec:tau_lh}
For what concerns the $\tau \to \ell h^0$ channel, pseudoscalar ($\pi^0$, $\eta$, $\eta'$ and $K_S^0$) and 
vector bosons ($\rho^0$, $\phi$, $\omega$, $K^*_0$) are considered in the final states, and reconstructed in different
final state, for a total of 17 decay channels. The best limits currently available are summarized in Table~\ref{tab:tau_lh}. 

\begin{table}[h]
\begin{center}
\caption{Best available limits on $\tau \to \ell h^0$ decays.}
\begin{tabular}{|l|c|c|}
\hline 
\textbf{Channel} & \textbf{90\% C.L.} & \textbf{Ref.} \\
 & \textbf{Upper Limit [$\times 10^{-8}$]}  & \\
\hline 
$\tau^- \to \mu^- \rho^0$ & 1.2 & Belle~\cite{belle_tau_lV} \\ 
\hline
$\tau^- \to e^- \rho^0$ & 1.8 & Belle~\cite{belle_tau_lV} \\ 
\hline
$\tau^- \to \mu^- \phi$ & 8.4 & Belle~\cite{belle_tau_lV} \\ 
\hline
$\tau^- \to e^- \phi$ & 3.1 & Belle~\cite{belle_tau_lV} \\ 
\hline
$\tau^- \to \mu^- \omega$ & 4.7 & Belle~\cite{belle_tau_lV} \\ 
\hline
$\tau^- \to e^- \omega$ & 4.8 & Belle~\cite{belle_tau_lV} \\ 
\hline
$\tau^- \to \mu^- K^{*0}$ & 7.2 & Belle~\cite{belle_tau_lV} \\ 
\hline
$\tau^- \to e^- K^{*0}$ & 3.2 & Belle~\cite{belle_tau_lV} \\ 
\hline
$\tau^- \to \mu^- \overline K^{*0}$ & 7.0 & Belle~\cite{belle_tau_lV} \\ 
\hline
$\tau^- \to e^- \overline K^{*0}$ & 3.4 & Belle~\cite{belle_tau_lV} \\ 
\hline
$\tau^- \to \mu^- \eta$ & 3.8 & Belle~\cite{belle_tau_lP} \\ 
\hline
$\tau^- \to e^- \eta$ & 3.6 & Belle~\cite{belle_tau_lP} \\ 
\hline
$\tau^- \to \mu^- \pi^0$ & 2.2 & Belle~\cite{belle_tau_lP} \\ 
\hline
$\tau^- \to \mu^- K_S^0$ & 3.3 & BaBar~\cite{babar_tau_lKS} \\ 
\hline
$\tau^- \to e^- K_S^0$ & 4.0 & BaBar~\cite{babar_tau_lKS} \\ 
\hline
\end{tabular}
\label{tab:tau_lh}
\end{center}
\end{table}

The analysis technique adopted for these decays is very similar to the one already used for the $\tau \to e \gamma$ and
$\tau \to 3\ell$ searches. At first, one selects events with two or four charged tracks, according to 
the final state under study. Then, working in the center of mass (CM) frame, one defines the \emph{thrust axis} 
of the event as the axis along which the sum of the projections of charged track and neutral candidate momenta 
is maximum. A plane perpendicular to the thrust axis divides the event into two hemispheres: one of them (the \emph{tag side}) is
required to contain only one charged track (identifying the decay of one $\tau$ into a 1-prong decay channel);
the other hemisphere has to contain the signal signature (including some particle identification requirements, if needed), 
no other track, and no other photon with energy above a given threshold (typically 100~MeV).
According to the number of tracks in the signal hemisphere, the event is said to have a 1-1 or a 1-3 topology. 
Once the event is selected in this way, the main background contributions come from 
$e^+e^- \to \tau^+ \tau^-$ and $e^+e^- \to \mu^+ \mu^-$ with initial state radiation, 
two-photon events $e^+e^- \to \gamma^* \gamma^* e^+ e^-$ and continuum 
$e^+e^- \to q \overline q$ ($q = u,\,c,\,s,\,d$) events. Moreover, some specific physical backgrounds can 
be relevant in a given channel. In the $\tau \to \ell \omega$ search, for instance, a significant 
contribution comes from $\tau \to \pi \omega \nu$, with the pion misidentified as a lepton.

The signal is separated from backgrounds by asking for the reconstructed mass of the signal 
$\tau$, $M_\tau$, to be around the nominal $\tau$ mass and the missing energy in the signal side, 
$\Delta E = E_\tau^{CM} - E_{beam}^{CM}$, to be around zero. Either these requirements are imposed 
for the events to be selected, and the number of surviving events is compared to the expected
background yield, or a likelihood for these observables is built, and used to set a limit on the signal
yield. 




From the theoretical point of view, setting limits on the $\tau \to \ell h$ decay rates is found to be
very effecting in constraining some specific NP model. For instance, it has been shown~\cite{THDM-III}
that the $\tau\mu$ Yukawa couplings in Two-Higgs-Doublet models of Type III are strongly constrained
by the results shown in Table~\ref{tab:tau_lh}.

\subsection{Searches for $\tau \to \ell h h'$ ($h = \pi^\pm$, $K^\pm$)}
Searches for $\tau \to \ell h h'$, with $h = \pi^\pm$, $K^\pm$, are performed with the same technique outlined in Sec.~\ref{sec:tau_lh}.
Events with 1-3 topology are selected, and 14 different modes are searched for. Eight of them ($\tau^- \to \ell^- h^+ h'^-$) 
violate the lepton flavor but conserve the global lepton number, the other six ($\tau^- \to \ell^+ h^- h'^-$) also imply LNV.

Relevant backgrounds arise from specific physical processes. In the $\tau \to \mu K \pi$ channel, pion-to-muon
and pion-to-kaon misidentification produces a significant background coming from $\tau \to \pi \pi \pi \nu$. In order to 
reduce this contribution, the invariant mass of the three charged tracks in the signal side, when they are all assigned a pion 
mass hypothesis, is required to be above 1.52 $\mathrm{GeV}/c^2$



The best limits for the $\tau \to \ell h h'$ decay modes, obtained by the Belle Collaboration~\cite{belle_tau_lhh}, are
reported in Table~\ref{tab:tau_lhh}. 
\begin{table}[h]
\begin{center}
\caption{Best available limits on $\tau \to \ell h h'$ decays~\cite{belle_tau_lhh}.}
\begin{tabular}{|l|c|}
\hline 
\textbf{Channel} & \textbf{90\% C.L.} \\
 & \textbf{Upper Limit [$\times 10^{-8}$]} \\
\hline 
$\tau^- \to \mu^- \pi^+ \pi^-$ & 2.1 \\
\hline
$\tau^- \to \mu^+ \pi^- \pi^-$ & 3.9 \\
\hline
$\tau^- \to e^- \pi^+ \pi^-$ & 2.3 \\
\hline
$\tau^- \to e^+ \pi^- \pi^-$ & 2.0 \\
\hline
$\tau^- \to \mu^- K^+ K^-$ & 4.4 \\
\hline
$\tau^- \to \mu^+ K^- K^-$ & 4.7 \\
\hline
$\tau^- \to e^- K^+ K^-$ & 3.4 \\
\hline
$\tau^- \to e^+ K^- K^-$ & 3.3 \\
\hline
$\tau^- \to \mu^- \pi^+ K^-$ & 8.6 \\
\hline
$\tau^- \to e^- \pi^+ K^-$ & 3.7 \\
\hline
$\tau^- \to \mu^- K^+ \pi^-$ & 4.5 \\
\hline
$\tau^- \to e^- K^+ \pi^-$ & 3.1 \\
\hline
$\tau^- \to \mu^+ K^- \pi^-$ & 4.8 \\
\hline
$\tau^- \to e^+ K^- \pi^-$ & 3.2 \\
\hline
\end{tabular}
\label{tab:tau_lhh}
\end{center}
\end{table}

\subsection{Searches for $\tau \to \Lambda h$ ($h = \pi^\pm$, $K^\pm$)}
There is a specific theoretical interest for the LNV decay modes $\tau \to \Lambda h$, ($h = \pi^\pm$, $K^\pm$), 
in NP models with higher generations~\cite{Hou}. These decays have been searched for by the Belle Collaboration, 
considering two modes that conserve the difference $B-L$ of the baryon number and the global lepton number ($\tau^- \to \overline \Lambda h^-$)
and two modes that also violate it ($\tau^- \to \overline \Lambda h^-$).

A 1-3 topology is used also for these modes, by reconstructing the $\Lambda \to p \pi$ decay modes. Two important background
sources require a specific treatment. At first, in $e^+e^- \to \tau^+ \tau^-$ decays, one of the two pions can decay into $K_S^0 \pi$, and
a $K_S^0 \to \pi^+ \pi^-$ decay can fake a $\Lambda$ if a pion is misidentified as a proton. This background can be reduced by
rejecting events where the two tracks forming the $\Lambda$ candidate, once a pion mass is assigned to both of them,
have an invariant mass that is consistent with a $K_S^0$ hypothesis. Second, continuum $e^+ e^- \to q \overline q$ events can 
produce a $\Lambda$ in association with charged and low-momentum neutral pions. In this case, baryon number conservation implies
a proton on the tag side, that can be vetoed to reduce this kind of contamination.

As reported in~\cite{belle_tau_Lambdah}, Belle set the 90\% C.L. upper limits $BR(\tau^- \to \overline \Lambda \tau^- ) < 2.8 \times 10^{-8}$
and $BR(\tau^- \to \overline \Lambda K^- ) < 3.1 \times 10^{-8}$ for the $(B-L)$-conserving modes, and 
$BR(\tau^- \to \Lambda \tau^- ) < 3.0 \times 10^{-8}$ and $BR(\tau^- \to \Lambda K^- ) < 4.2 \times 10^{-8}$ for the $(B-L)$-violating modes.

\subsection{Searches for $\tau$ LFV at LHC}
At LHC, a large amount of $\tau$ leptons is produced in the decays of $B_{(s)}$ and $D_{s}$ mesons. At LHC$b$, the cross section for $\tau$
production is around 80~nb. As a result, LHC experiments can productively search for LFV in $\tau$ decays, at least for these modes with only charged tracks
in the final state. The first limit set by LHC$b$, $BR(\tau \to 3\mu) < 6.3 \time 10^{-8}$ at 90\% C.L. has been first presented at 
this conference. It is impressive how LHC$b$ already starts to be competitive with $B$-Factories in this field. 
Searches for $\tau \to \mu \phi$ and $\tau \to \mu h h'$ could be also feasible in the LHC environment.

\section{LFV IN THE MUON SECTOR}

\subsection{The MEG Experiment} 
LFV searches in the muon sector require dedicated muon beams with very high intensity and purity. At present, the most intense continuous muon beam in the
world is available at the Paul Scherrer Institut (PSI, Switzerland), and can provide up to $10^{8}$ muons per second. The MEG experiment, that searches for the
$\mu \to e \gamma$ decay with a goal sensitivity of a few $10^{-13}$, is operated at this facility.

\begin{figure}[t]
\centering
\vspace{-4cm}
\includegraphics[width=90mm]{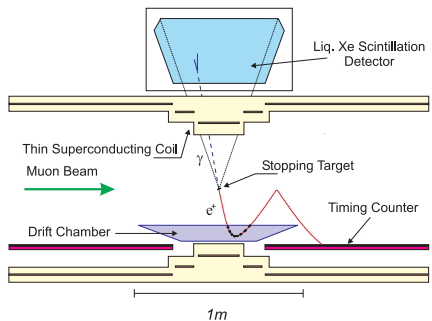}
\vspace{-3cm}
\caption{The MEG experiment} \label{fig:MEG}
\end{figure}

In the MEG experiment, depicted in Figure~\ref{fig:MEG}, positive muons are stopped in a thin polyethylene target and their decay products 
are analyzed by a Liquid Xenon calorimeter (XEC) and
a spectrometer, composed of a drift chamber (DCH) system and a scintillating bar hodoscope (Timing Counter, TC) for positron timing and trigger, inside a
superconducting magnet. The signal signature is provided by a photon and a positron, emitted simultaneously and back-to-back, with an energy 
$m_\mu c^2/2 \sim 52.8$~MeV.  (in the very good~  approximation of null positron mass). Due to the very high muon rate available at PSI, the dominant 
background is given by the accidental time coincidence of a positron, produced in the decay of a muon, and a photon, produced in the decay of another muon 
or by annihilation in flight of another positron. A subleading contribution comes from radiative muon decays (RMD) $\mu \to e \nu \overline \nu \gamma$.
Since the accidental background rate increases quadratically with the beam intensity, a muon rate of $3 \times 10^{7}$ muons per second is used. 
Indeed, given the present detector performances, there is no statistical advantage from an increase of the rate above this value, and only 
further difficulties would arise in the detector operation.

The XEC is composed by a 800 liter Liquid Xenon vessel, where the scintillation light produced by the photon shower is read out by 846 photomultiplier tubes.
Liquid Xenon is characterized by a fast response (45 ns for the scintillation light produced by photons) and a good light yield (75\% of the typical NaI(Tl) one), 
that allow to reach a time resolution of $\sim 80$~ps, an energy resolution of $\sim 1$~MeV at 52.8~MeV and a conversion point 
position resolution of $\sim 5$ mm. A redundant set of calibration and monitoring tools is available, including a Cockroft-Walton proton accelerator for
monitoring with low-energy photons from nuclear reactions~\cite{CW} and a pion beam for calibration with high-energy photons from the charge exchange
reaction $\pi^- p \to \pi^0 n$, $\pi^0 \to \gamma \gamma$. 

The DCH system is composed of 16 chambers, with two planes of 9 staggered drift cells each, whose wires are parallel to the beam axis. 
A mixture of Helium and Ethane in 50:50 volume proportions is used inside the chambers, and the system is immersed in a helium atmosphere, 
to reduce the multiple Coulomb scattering. Chambers are closed by aluminized kapton foils, which work as cathodes and are segmented to 
allow a measurement of the longitudinal coordinate with the Vernier pattern method~\cite{vernier}. A total of only $\sim 10^{-3}$ radiation lengths
is encountered by a positron through its path toward the TC, that is composed of 30 bars of plastic scintillator, divided in two sectors, for tracks
going upstream and downstream with respect to the beam. The DCH spectrometer provides a core momentum resolution of $\sim 320$~MeV, angular resolutions
at the level of $9/8$~mrad in $\phi$/$\theta$, and a vertex resolution of $\sim 5/3$~mm in $Y$/$Z$. The TC bars have a timing resolution of 60 to 80~ps, to
be combined with a flight length resolution of $\sim 70$~ps from the target to the TC. The global $e\gamma$ relative time resolution turn out to be
$\sim 130$~ps.

\subsection{MEG Results from 2009 and 2010 Runs}

\begin{figure*}[t]
\centering
\includegraphics[width=150mm]{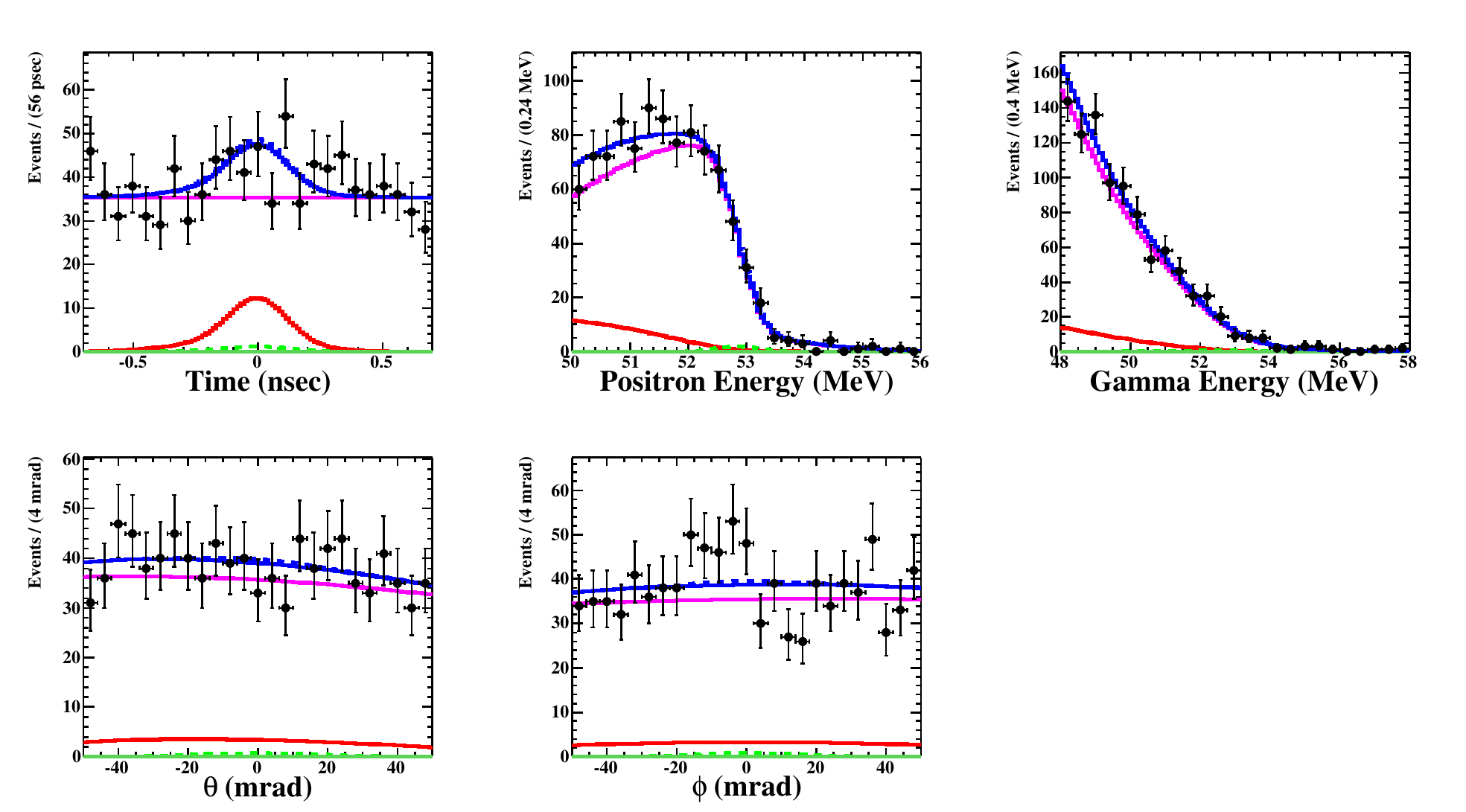}
\caption{Maximum likelihood fit to the event distribution in the signal region. The green, red and magenta lines represent the fitted signal, RMD 
and accidental background contributions, respectively, while the blue line shows the total PDF.} \label{fig:fit}
\end{figure*}

MEG results from data collected in 2009 and 2010 are presented here, for a total of $\sim 17.5 \times 10^{13}$ muons stopped at the target. Data were
analyzed with a likelihood technique, in order to fully exploit the discriminating power of the photon energy $E_\gamma$, the positron momentum $E_e$, the
relative polar angles $\theta_{e\gamma}$, $\phi_{e\gamma}$ and the relative time $T_{e\gamma}$. The probability density function (PDF) for the accidental
background was obtained from data, using only events out of a signal region in the $(E_\gamma,T_{e\gamma})$ plane. The same control sample, 
along with data from calibrations, was used to extract the signal and RMD PDFs. In particular, the photon energy resolution is extracted from the
$\pi^-$ charge exchange data; the photon conversion point resolution from special runs with collimators in front of the detector;
the positron momentum resolution from a fit of the spectrum for the normal muon decay; vertex and angle resolutions 
by comparing different track segments in events where the positron makes multiple turns inside the spectrometer; the timing resolution from
RMDs. The multiple turn technique also allows to study several correlations that arise in the positron reconstruction, due to the fact that the
decay vertex is defined as the intersection of the track with the target plane. Important correlations are present, in particular, between the reconstructed
$\phi_e$ angle and the reconstructed momentum and $\theta_e$ angle. For the same reason, the $\phi_e$ resolution shows a dependence on $\phi_e$, with a
minimum for $\phi_e = 0$. These and all other relevant correlations are included in the PDFs for the likelihood analysis. 

The sensitivity of the experiment is then determined by generating a set of pseudo-experiments according to the PDFs. The confidence intervals for 
each pseudo-experiment are determined using a fully frequentistic approach inspired to the Feldman-Cousins technique~\cite{FC}, with a maximum 
likelihood ratio ordering~\cite{PDG} to treat the uncertainty on the number of background events. The expected Upper Limit (UL) at 90\% confidence 
level (CL) is $1.6 \times 10^{-12}$ from the combination of 2009 and 2010 data, and $3.3 \times 10^{-12}$ ($2.2 \times 10^{-12}$) from 2009 (2010) alone. 
The combined sensitivity is 7.5 times lower than the previous best limit on $\mu \to e \gamma$, set by the MEGA collaboration 
at $1.2 \times 10^{-11}$~\cite{MEGA}. 

\begin{figure}[h]
\centering
\includegraphics[width=40mm]{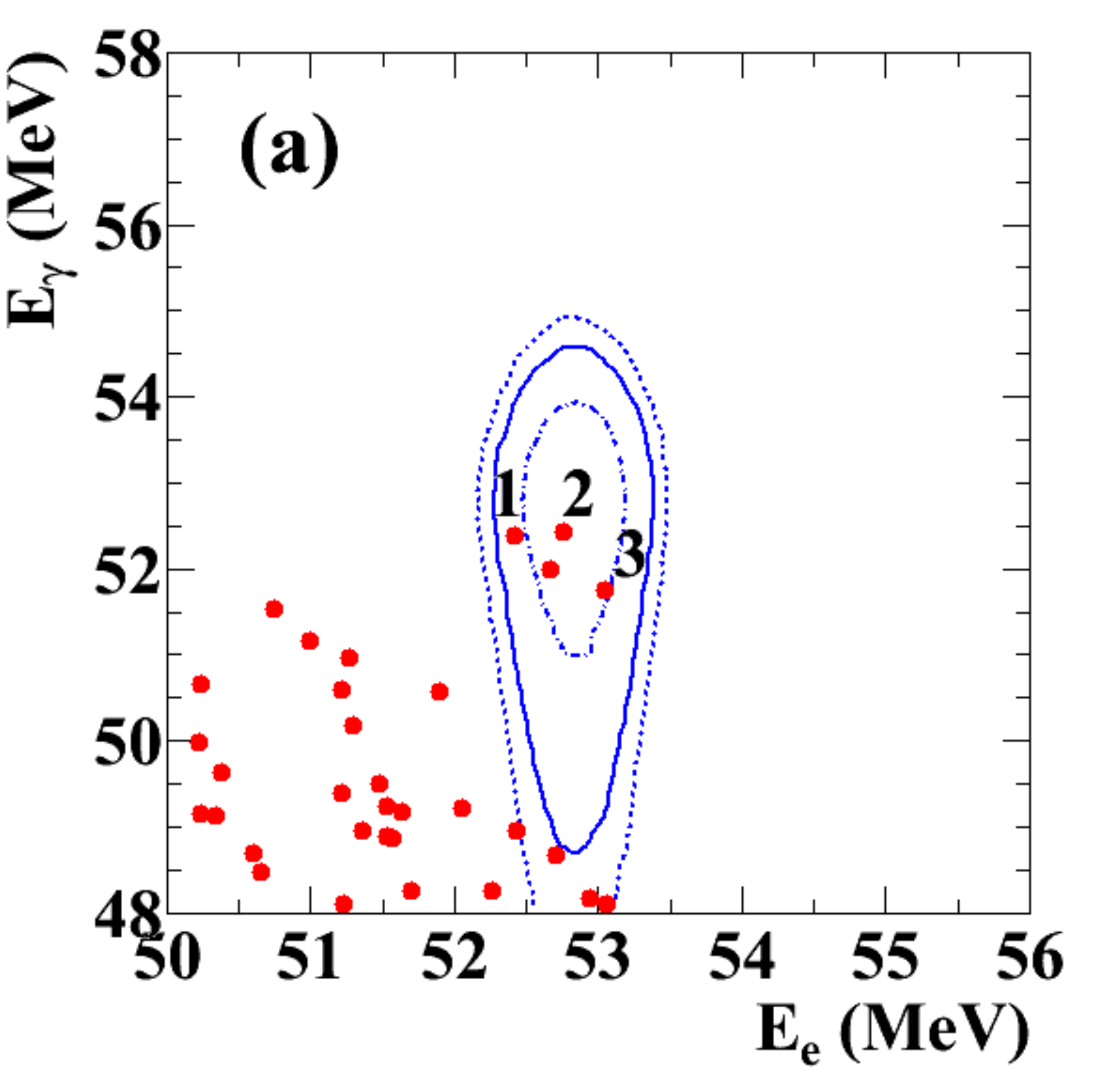}
\includegraphics[width=40mm]{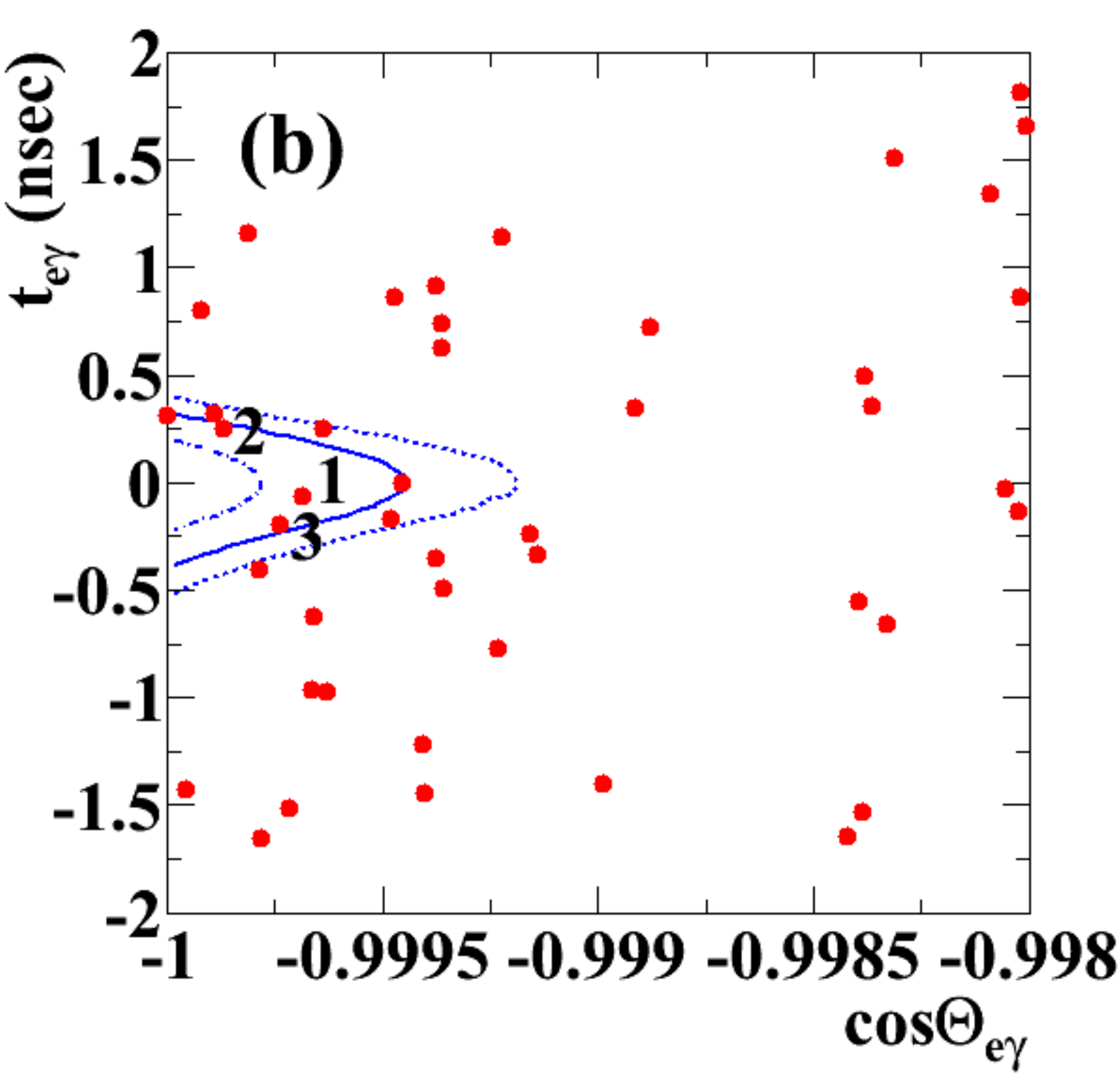}
\includegraphics[width=40mm]{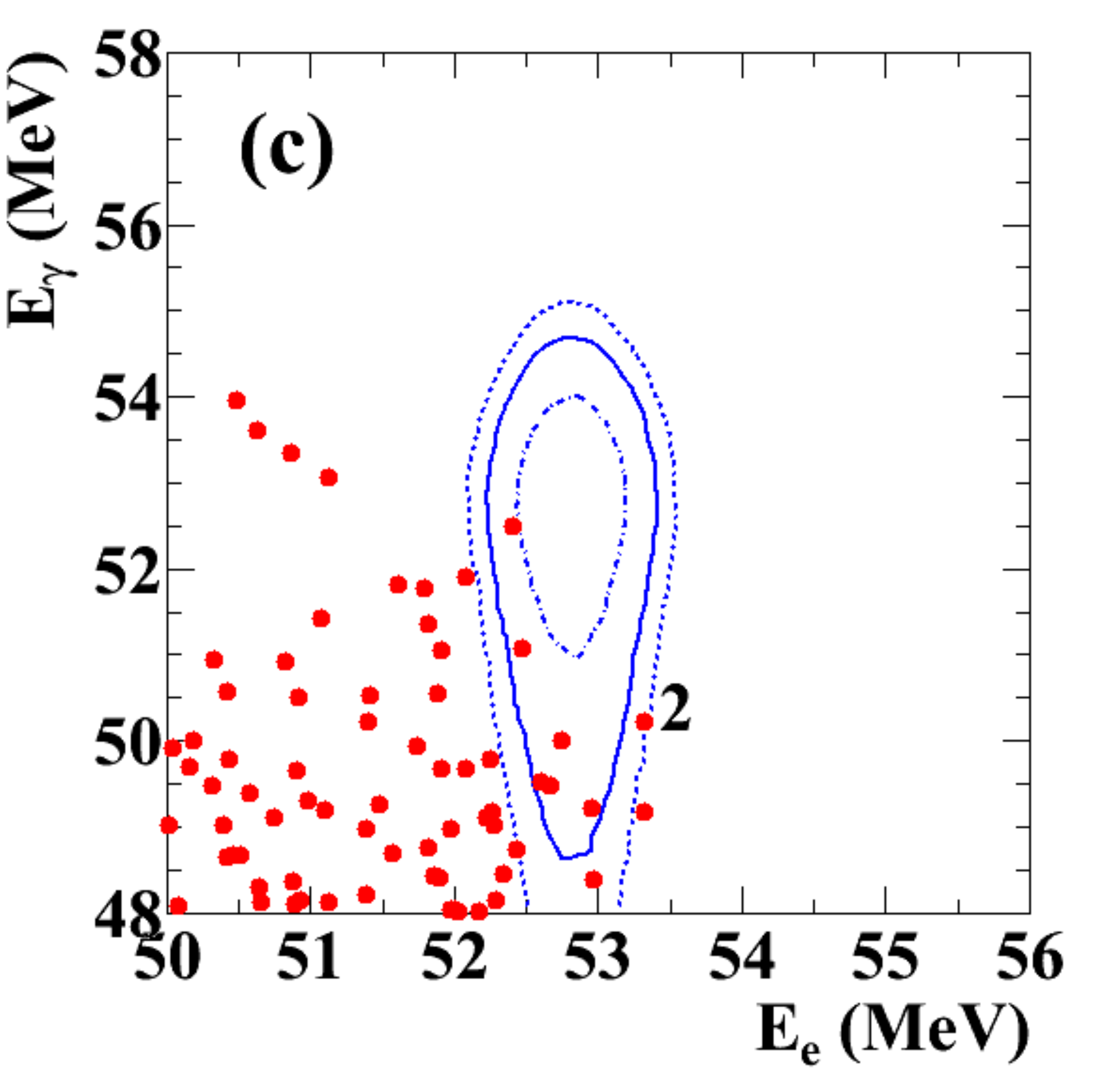}
\includegraphics[width=40mm]{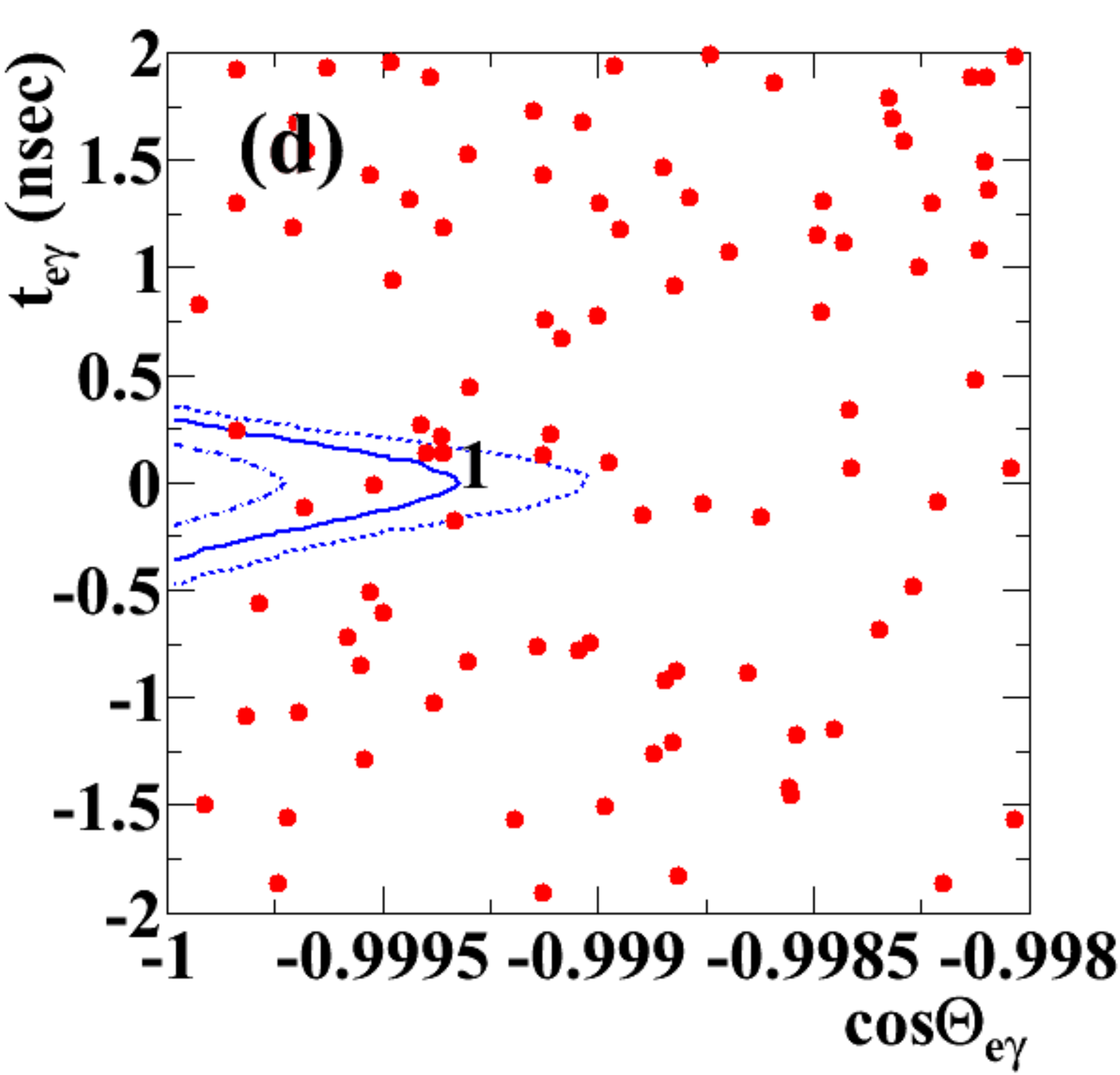}
\caption{Event distribution projected in the $(E_e,E_\gamma)$ plane (left) and in the $(T_{e,\gamma},cos(\Theta_{e,\gamma}))$ plane (right), being $\Theta_{e\gamma}$ 
the angle between the positron and the photon. Top: 2009 data. Bottom: 2010 data. For each projection, a 90\% signal efficiency selection is applied on the
other.} \label{fig:events}
\end{figure}

Once  the analysis was fully developed, it has been applied to the data inside the signal region. The result of the fit is shown in Figure~\ref{fig:fit}
and the maximum likelihood estimate for the signal, RMD and accidental background yields is $N_{sig} = -0.5^{+7.9}_{-4.7}$, $N_{RMD} = 76.5 \pm 12$, 
and $N_{acc} = 882 \pm 22$, respectively. Some 2-dimensional event distributions are shown in Figure~\ref{fig:events}. The frequentistic approach described
above is used to set an UL $BR(\mu \to e \gamma) < 2.4 \times 10^{-12}$ at 90\% CL~\cite{meg}, that improves by a factor of 5 the MEGA limit and by a factor
of 12 the previous MEG limit~\cite{meg2008}.

\subsection{MEG Status}

The MEG experiment collected new data on 2011, corresponding to $\sim 19 \times 10^{13}$ stopped muons, corresponding
to 1.1 times the statistics used for the latest published result~\cite{meg}. The analysis of these data is on going, with significant
improvements in the reconstruction algorithms for both the positron and the photon side. A sensitivity at the level of $10^{-12}$ is expected. 
A further data taking is planned on 2012, and should allow to collect a similar statistics. A sensitivity of $\sim 7 \times 10^{-13}$ could be reached.
According to the sensitivity projections, the upper limit scales faster than the square root of the statistics, but more slowly than linearly, indicating that
a relevant background starts to arise. Hence, a further significant progress in a reasonable amount of time will be only possible with a major upgrade
of the experiment, apt to reduce the influence of the background by improving significantly the present resolutions.

\section{FUTURE PERSPECTIVES}

Several projects aim to improve the present limits on LFV and LNV both in the $\tau$ and muon sector. The two Super B-Factories under study,
Super-KEKB (KEK, Japan)~\cite{SuperKEKB} and SuperB (Cabibbo-Lab, Italy)~\cite{SuperB} should allow to set limits at a level of $10^{-9}$ in the tau sector. 
At the same time, two experiments should reach a sensitivity below $10^{-16}$ in the search for the $\mu \to e$ conversion process in the interaction of
negative muons with nuclei by which they are captured: the Mu2e experiment~\cite{mu2e} at Fermilab (USA) and the COMET experiment~\cite{comet} at 
J-PARC (Japan).

Two R\&D plans are on going at PSI. The first one, named Mu3e~\cite{mu3e}, proposes the search for $\mu^+ \to e^+ e^+ e^-$, 
by means of a magnetic spectrometer instrumented with MAPS silicon detectors. According to its design performances, is should allow to 
reach a sensitivity around $10^{-16}$, that would be competitive with the MEG limits on $\mu \to e \gamma$ in the models that predict both decays, 
and would be sensitive to NP contributions not probed by $\mu \to e \gamma$. The second plan is an upgrade of the MEG detector, including a new 
tracking system for the positron spectrometer, new photon detectors to replace the PMTs of the XEC and a new design of the TC. An upper limit of 
$5 \times 10^{-14}$ could be reached with this upgrade, making it competitive with the first phase of the $\mu \to e$ conversion experiments. 

\bigskip 
\begin{acknowledgments}
The author wish to thank G.~Isidori and J.~Jones-Perez for providing customized plots for this talk.
\end{acknowledgments}

\bigskip 

\end{document}